\documentclass[9pt]{article}
\usepackage{spconf}
\usepackage{amsmath}
\usepackage{graphicx}
\usepackage{tabularx}
\usepackage{lipsum}
\usepackage{booktabs}
\usepackage{multirow}
\usepackage{hhline}
\usepackage{tabularx,booktabs}
\usepackage{float}
\usepackage{epsfig}
\usepackage{latexsym}
\usepackage{psfrag}
\usepackage{pstricks}
\usepackage{dsfont}
\usepackage{cancel}
\usepackage{epstopdf}
\usepackage{bigints}
\usepackage{cite}
\usepackage{url}
\usepackage[normalem]{ulem}
\usepackage{multirow}
\usepackage{amsfonts}
\usepackage{amssymb}
\usepackage{xcolor,colortbl}
\usepackage{array}
\newcolumntype{P}[1]{>{\centering\arraybackslash}p{#1}}
\newcolumntype{M}[1]{>{\centering\arraybackslash}m{#1}}

\definecolor{lightmauve}{rgb}{0.86, 0.82, 1.0}
\definecolor{pastelblue}{rgb}{0.68, 0.78, 0.81}
\definecolor{timberwolf}{rgb}{0.86, 0.84, 0.82}
\definecolor{anti-flashwhite}{rgb}{0.95, 0.95, 0.96}
\definecolor{grannysmithapple}{rgb}{0.66, 0.89, 0.63}
\definecolor{melon}{rgb}{0.99, 0.74, 0.71}
\definecolor{melon1}{rgb}{1, 0.88, 0.87}
\definecolor{melon2}{rgb}{1, 0.82, 0.80}

\newcolumntype{a}{>{\columncolor{timberwolf}} c} 

\title{Effectiveness of Random Deep Feature Selection for securing image manipulation detectors against adversarial examples}
\name{$^*$M.Barni$^{\dagger}$, E.Nowroozi$^{\dagger}$, B.Tondi$^{\dagger}$, B.Zhang$^{\#}$
 \thanks{This work has been partially supported by a research sponsored by DARPA and Air Force Research Laboratory (AFRL) under agreement number FA8750-16-2-0173. The U.S. Government is authorised to reproduce and distribute reprints for Governmental purposes notwithstanding any copyright notation thereon. The views and conclusions contained herein are those of the authors and should not be interpreted as necessarily representing the official policies or endorsements, either expressed or implied, of DARPA and Air Force Research Laboratory (AFRL) or the U.S. Government.\newline * The list of authors is provided in alphabetic order.}\vspace{-0.2cm}
\address{$^{\dagger}$ Department of Information Engineering and Mathematics, University of Siena, Italy.\\
$^{\#}$ School of Cyber Engineering, Xidian University, China.}
}

\vspace{-2cm}

\begin{document}
\ninept

\maketitle
\begin{abstract}
We investigate if the random feature selection approach proposed in \cite{chen2019secure} to improve the robustness of forensic detectors to targeted attacks, can be extended to detectors based on deep learning features. In particular, we study the transferability of adversarial examples targeting an original CNN image manipulation detector to other detectors (a fully connected neural network and a linear SVM) that rely on a random subset of the features extracted from the flatten layer of the original network. The results we got by considering three image manipulation detection tasks (resizing, median filtering and adaptive histogram equalization), two original network architectures and three classes of attacks, show that feature randomization helps to hinder attack transferability, even if, in some cases, simply changing the architecture of the detector, or even retraining the detector is enough to prevent the transferability of the attacks.
\end{abstract}
%

\section{INTRODUCTION}
\label{sec:intro}

The development of secure image forensic tools  achieving  good performance even in the presence of an adversary aiming at impeding the analysis is not an easy task, given the weakness of the traces the forensic analysis relies on \cite{Gloe07}.
When the attacker has a perfect knowledge about the forensic algorithm (white-box scenario) or enough information about it,
powerful Counter-Forensic (CF) techniques can be devised, by introducing a limited distortion into the attacked image.
This is even more the case with Deep Learning (DL)-based forensics,  due to the vulnerability of DL techniques to adversarial examples \cite{goodfellow2014explaining}, small quasi-imperceptible perturbations of the input images causing  incorrect results \cite{BestaAdv17, VerdoAdv18, barni2018adversarial}.

A possibility to improve the general robustness against CF, without specializing the forensic algorithm against a particular CF tool (as it is the case with adversary-aware approaches, see for instance \cite{barni2017eusipco} for machine learning-based forensics and \cite{goodfellow2014explaining} for general DL applications), is to resort to randomization strategies \cite{barni2018adversarial}. In fact, many randomization approaches have been considered addressing standard machine learning tools and, more recently, DL architectures. In \cite{chen2019secure}, the authors propose to randomize the selection of the feature space according to a secret key to prevent the attacker from gaining full knowledge about the system. In this way, the analyst {\em exits} the white-box scenario \cite{yuan2019adversarial} thus decreasing the success rate of the attack.
The effectiveness of Random Feature Selection (RFS) has been  proven in \cite{chen2019secure}
both theoretically, under simplifying assumptions, and in practice, where it is experimentally validated by focusing on image manipulation detection and the SPAM feature set \cite{pevny2010steganalysis}.
With regard to DL techniques, most of the methods proposed so far focus on test time randomization
\cite{xie2017mitigating}, where  the input layer of the classifier is randomized at test time.
A multi-channel architecture,
where each channel introduces its own randomization in
a special transformed domain based on a secret key, has recently been proposed in \cite{taran2019defending}.

In this paper, we extend the random feature selection approach described in \cite{chen2019secure} to the case of CNN-based  detection, where the features are extracted by a convolutional neural network, to see if and up to which extent the approach can be used to combat adversarial examples. In the following, we refer to the new scheme as Random Deep Feature Selection (RDFS). To perform the classification based on the randomly selected deep features, we consider two types of classifiers, a Fully Connected network and a linear SVM. With regard to the FC network, it corresponds to a retrained version of the last part (the FC layers) of the original CNN targeted by the attack.

To be effective, RDFS should improve the security of CNN-based detectors against adversarial examples, at the expense of a negligible loss of performance in the absence of attacks.
The experiments we carried out on several image manipulation detection tasks, considering two state-of-the-art CNNs, reveal that the dangerousness of adversarial examples can indeed be mitigated
by the proposed RDFS scheme. However, the degree of effectiveness of  RDFS depends on the detection task, the kind of attack and the network.
In fact, in several cases,
the mismatch between the original classifier targeted by the attack and the one used for classification (in our case an SVM or a retrained FC classifier)
decreases by itself the attack success rate, thus making feature randomization unnecessary.


The rest of this paper is organised as follows. In Section \ref{sec.RDFS} we present the general RDFS scheme. Then, in Section  \ref{sec.METH}, we describe the methodology we followed for our experiments, where we applied the RDFS scheme to several image manipulation detection tasks. The results we got are reported and discussed in Section \ref{sec.EXP}. Finally, in Section \ref{sec.CONC}, we discuss possible extensions of this work.

\section{Random Deep Feature Selection (RDFS) for Secure Image Classification}
\label{sec.RDFS}



\begin{figure}[t!]
	\centering	
		\includegraphics[width=0.90\columnwidth]{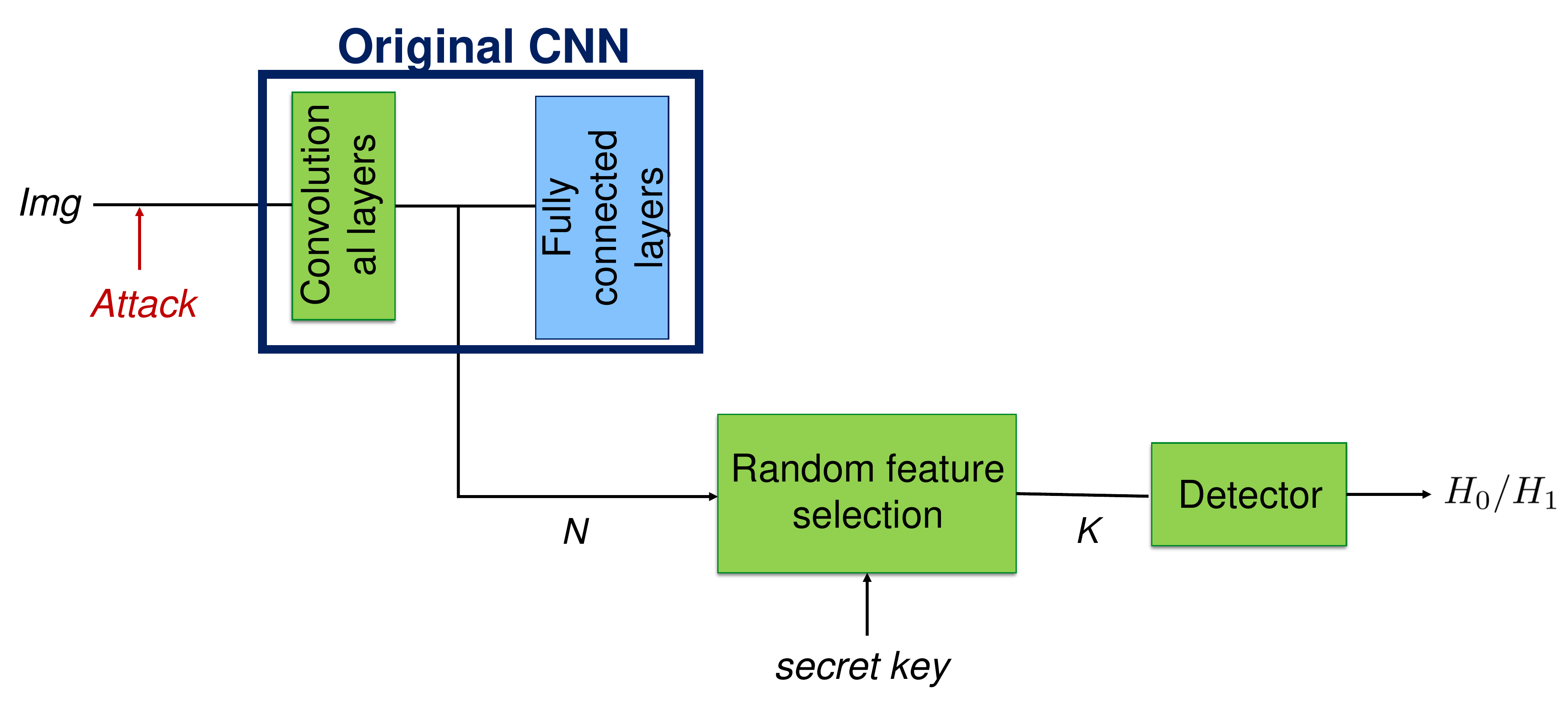}	
\vspace{-0.2cm}
	\caption{Scheme of the proposed RDFS detector.}
	\label{fig:secmodel}
\vspace{-0.5cm}
\end{figure}

As we said, our goal is to extend the random feature selection method developed in  \cite{chen2019secure} for
model-based and standard ML-based detectors based on statistical and handcrafted features, to the case of CNN-based forensic detectors.
The security model considered in this paper is depicted in Fig. \ref{fig:secmodel}. Given an original CNN detector, the CNN is only used as feature extractor. Let $N$ be the dimensionality of the set of features. Then, $K$ features ($K < N$) are randomly selected among the $N$ features, according to a secret key. The reduced set of features obtained in this way is used to train another detector, for instance, a Neural Network or an SVM classifier.
Obviously, the same scheme is applied  during both training and testing, with the same secret key.
Without loss of generality, we let $H_0$ be the hypothesis that the image is original,  and $H_1$  the hypothesis that the image has been tampered with.
The adversarial attack is carried out in the pixel domain, as shown in Fig. \ref{fig:secmodel}.
We assume that the attacker does not know the existence of the randomization strategy and then he targets the original CNN classifier (hence implementing a so called {\em vanilla attack}) \cite{athalye2018obfuscated,xie2017mitigating}.
%
Also, we assume that the attacker wants to pass off a manipulated image as an original one, i.e., to induce the network to decide in favor of $H_0$ when $H_1$ holds, 
causing a false negative error, while he is not interested in attacking in the opposite direction.
At the same time, the attacker wants to minimize the distortion introduced in the image as a consequence of the attack.

As mentioned in the introduction, the random feature selection enforces a limited knowledge scenario for the attack.
In particular,  the attacker is not aware of the randomization defence mechanism (and, even if he is, he does not know the subset of features used by the detector and their number $K$). Moreover, he has only a partial knowledge of the RDFS architecture of the detector.
By following a terminology of DL \cite{yuan2019adversarial},  we can say that, due to  random feature selection, the attack is no more carried out in a white-box setting; the random feature selection induces a so called {\em semi white-box} scenario for the attack.
%
The exact amount of knowledge available to the attacker depends on the specific RDFS scheme considered, that is, on the  specific architecture of the detector. We considered two different scenarios: the case of a Fully-Connected (FC) network detector, and the case of a linear SVM detector.

\subsection{RDFS detection based on a Fully-Connected (FC) network}

In this case, the RDFS detector is implemented by retraining the FC layers of the original CNN.  Given the selected random feature set of dimensionality $K$ ($K < N$), the same FC structure of the original network is  re-trained considering only the $K$ input nodes.
The number of layers depends then on the original CNN architecture.
Therefore, when the full feature set is considered for the detection ($K =N$), there is no mismatch in the  architecture between this FC network and the classification architecture of the original CNN targeted by the attack.
However, even in this case, the attack is not completely white-box, since, in the setup considered for our experiments, a different training set (more precisely, a subset of the original set) is used to train the detector.

\subsection{RDFS detection based on SVM}

In this case, an SVM architecture is considered to implement the detector.
Then, in contrast to the previous case, the detector is different from the classification structure of the original CNN, even when the full feature set is considered ($K =N$).
The amount of attack knowledge is less than in the previous case, since the attacker does not know the architecture of the detector, in addition to the training data.


\section{APPLICATION TO MANIPULATION DETECTION}
\label{sec.METH}


%


In this paper, we are interested in evaluating the performance of the RDFS approach and assess the factors that affect its effectiveness 
for image manipulation detection applications. Specifically, we run several experiments by considering  three different manipulation detection tasks. For each of them, two different state-of-the-art CNN architectures were considered. The security of the RDFS approach was assessed by considering three different types of attacks.

\subsection{Original CNN Detectors}

We considered three different  detection tasks, namely the detection of image resizing (downsampling, by a 0.8 factor),  median filtering (by a 5 $\times$ 5 window), and adaptive histogram equalization (AHE), which applies
contrast enhancement on a local basis (the Contrast-Limited implementation of AHE, namely, CL-AHE, was
considered).
Concerning the network architectures, we considered the network in \cite{bayar2016deep} (recently extended in \cite{BayarCNNUnivTIFS}), referred to as BayarNet2016, and the one in \cite{BCNT18}, referred to as BarniNet2018. BayarNet2016 was originally proposed for the detection of some standard manipulation detection tasks and consists of 3 convolutional layers followed by batch normalization, 3 max-pooling layers, and 2 FC layers. As a main feature of BayarNet2016, the filters of the first layer (with $5 \times 5$ receptive field) are constrained  by enforcing
a high-pass nature of the filters, and then residual-based features are extracted.
BarniNet2018 was proposed for the more difficult task of generic contrast adjustment detection (the accuracy achieved by BayarNet2016 on this task is quite low). The main features of BarniNet2018, making it significantly different from  BayarNet2016, are
that it is pretty deep (9 convolutional layers instead of 3), and    no pre-filtering is applied by constraining the filters of the first layer (then the features are automatically learned by the networks from the image pixels).
We refer to \cite{bayar2016deep} and \cite{BCNT18} for the  details on the network structures.

\subsection{Reduced (Deep Feature) Detectors}

For each manipulation task, we built our reduced feature detectors as detailed in the following.
For a given choice of the random selection, the $K$ features were extracted from the flatten layers of the CNNs.
Then, we trained the original FC architecture of BayarNet2016 and BarniNet2018 with $K$ input nodes, and  a linear SVM classifier fed with the $K$ input features. Both the FC networks and the SVM were trained on a subset of images of the training set used to train the original networks.
Regarding the structure of the two FC networks,  we have 2 layers with 4096 hidden nodes each
for BayarNet2016, and only one layer consisting of 250 hidden nodes for BarniNet2018.
%
For the SVM, a Gaussian
kernel was adopted, and the kernel parameters were
determined by 5-fold cross-validation.

To measure the performance of the reduced set detectors, for a given size $K$ of the reduced feature set, the experiments were repeated (both training and testing)  50 times, each time with a different randomly chosen feature subset.

\enlargethispage{\baselineskip}



\subsection{Adversarial Attacks}

As we said,  the attack is targeted to the original CNN model.
In all the cases, {\em white-box} attacks are run against the original CNN trained models, i.e., assuming that the attacker knows everything about the model.
%
In our experiments, we considered three gradient-based, iterative, attacks.
Specifically, the adversarial examples were built by relying on the following algorithms:  the original box constrained L-BFGS, namely L-BFGS-B (BFGS for short) by Szegedy et al. \cite{Szegedy2013IntriguingPO}, the Fast Gradient Sign Method, namely FGSM, originally proposed in \cite{goodfellow2014explaining},
and the Projected Gradient Descent (PGD) attack,  originally proposed in 
\cite{madry2017towards}.
The Foolbox toolbox \cite{rauber2017foolbox} was used to implement the attacks.
L-BFGS  \cite{Szegedy2013IntriguingPO} looks for an approximately optimum solution of the optimization problem the attack has to solve, to find the minimal adversarial perturbation that makes the prediction change (standard gradient-descent algorithm).
%
For the FGSM attack, we considered the refined iterative version (I-FGSM) described in \cite{kurakin2016adversarial}.
At each iteration  $i + 1$, an adversarial
perturbation in obtained by computing the gradient
of the output with respect to the input image and considering its
sign multiplied by a (normalized) strength factor $\varepsilon$. Formally, given an image $X$,  $X_{i+1} = X_{i} + \varepsilon(\max(X_{i}) - \min(X_{i}))\cdot \text{sign}(\nabla_X J_{\theta}(X_{i},y))$, where $J_{\theta}(X,y)$  is the cross-entropy cost function of the neural network with parameters $\theta$, and $y$ is the ground truth class  of $X$.
The algorithm is applied iteratively until an adversarial image is obtained (that is, an image which is misclassified by the network), for a maximum number of steps $S$. Several values of $\varepsilon$ are considered, i.e. $\varepsilon \in E$; the value which minimizes the distortion of the final attacked image with respect to the original one is eventually selected.
%
PGD is about finding the perturbation that maximizes the loss function under some restrictions regarding the introduced $l^{\infty}$ distortion.
Specifically, at each iteration $i + 1$:  first the
image is updated (similarly to FGSM) as $X_{i+1} = X_{i} + \varepsilon (\max(X_{i}) - \min(X_{i})) \cdot \text{sign}(\nabla_X J_{\theta}(X_{i},y))$
for some $\varepsilon$ called step size;
then, the pixel values are
clipped to ensure that they remain in the $\alpha$-neighbourhood of the original image: accordingly, the image is refined by computing $[X_{i+1}]_{r,c} = \text{clip}([X_{i+1}]_{r,c}, \{-\alpha,+\alpha\})$ for pixel $(r,c)$.
Due to the clipping, this attack may result in a highly suboptimal solution in some cases.
A binary search can be performed over $\varepsilon$ and $\alpha$ to optimize the choice of the hyperparameters, using the input values only for the initialization.

\section{EXPERIMENTAL RESULTS AND DISCUSSION}
\label{sec.EXP}

In this section, we  first describe the experimental setup for our experiments, then we present and discuss the results of the tests we carried out in the various settings.


\subsection{Experimental setup} 
\label{sec.ExpSetup}

For our experiments, we considered  the 8156  uncompressed camera-native (.tiff) images from RAISE dataset \cite{RAISE8K} with size 4288 $\times$ 2848.
The images were split into training, validation, and test sets and then processed to create the images for the $H_1$  class, i.e., resizing, median, and adaptive histogram equalization (CL-AHE).  For all our experiments, the images were converted to gray-scale.

To build the original CNNs for the three detection tasks,   we considered 100.000 patches for training, 3000 and 10.000 for validation and testing respectively, per class, for BayarNet2016: for BarniNet2018, 500.000 patches were considered for training, 5000 and 10.000 for validation and testing respectively, per class. In order to use many images from the dataset and then enforce patch diversity, a maximum number of 100 patches were selected randomly from each image.  
For both networks, the input patch size was set to  $64\times 64$.
A number of  40 epochs was considered to train the BayarNet2016 models, while, by following \cite{BCNT18}, 4 epochs were considered for BarniNet2018.
%
For training the networks, we used the Adam solver with learning rate $10^{ - 4}$ and momentum 0.99. The batch size for training was set to 32.
The accuracies achieved by the trained models  in the absence of attacks are:
i)  BayarNet2016: $91.30\%$ for resizing,  $98.83\%$ for median filtering, and $90.45\%$ for CL-AHE; i)  BarniNet2018:
$95.05\%$ for resizing,
$99.73\%$ for median filtering,
and $ 98.30\%$ for CL-AHE.

In order to build the  RDFS detectors, both the FC networks and the SVM were trained on a subset of 20.000 patches per class selected from the original training set of images, and validated  on 1000  patches per class,   selected from the validation set used for the original CNN.
Regarding the training procedure for the reduced feature FC networks, we used the following setting: learning rate of the Adam solver set to $10^{ - 5}$ and the momentum to 0.99 for a maximum number of 50  epochs, with an early stop condition that ends training when the validation
loss changes less than $10^{-3}$ in the last 5 epochs (with a validation batch size of 100).
The number of reduced features $K$ considered in our experiments (for training and testing) for both the FC networks and the SVM is:  5, 10, 30, 50, 200, 400, 600, and the full feature case $K=N$.
The full feature set size  (size of the  flatten layer of the original CNN) is $N= 1728$ with BayarNet2016 and $N=3200$ with BarniNet2018.
When  BarniNet2018 is used as original CNN, the case $K = 600$ is not considered, to save time. For every value of $K$, in fact, we need to train 50 models (SVM and FC), one for every choice of the random set takes time. However, as confirmed by the results  with BayarNet2016,
the most interesting cases are those with lower values of $K$ (order of tens).
%
For the tests in the absence of attacks, we considered 4000 patches per class, taken from the original test set.

With regard to the attacks, for L-BFGS, we used the default attack parameters \cite{rauber2017foolbox}.
%
For FGSM, the number of steps $S$ is fixed to 10 (default), the best strength is searched in the range  $E  = [0: 0.001, 0.1]$.
%
PGD is applied considering the following setting:  $\varepsilon=0.05$, and $\alpha=0.3$, binary-search = 'True'\footnote{We refer to \cite{rauber2017foolbox} for the exact technical meaning of this parameter.}.,
%
%
The above setting does not work for the CL-AHE detection task (the adversarial image cannot be found in most of the cases); for that task, the following setting has been considered: $\varepsilon=0.025$, and $\alpha=0.01$, binary-search = 'False'.
%
%
%
As we said, we only applied the attack to images of the $H_1$ class.  In all the cases, the performance in the presence of attacks is evaluated on 500 adversarial examples, obtained by attacking a subset of the 4000 patches of the $H_1$ class.
We checked that, with the above setting, all the attacks are successful against the target original CNN, the success rate being in the range [0.98:1].
The average PSNR for the attacked samples is between 40 and 70dB (often above 60 dB), the exact value depending on the attack type, the target network and the detection task.


\subsection{Results of FC-based classification}

The results we have got for this case are reported in Table \ref{tab.FC_Bayar} and \ref{tab.FC_ICIP} for the case of BayarNet2016 and BarniNet2018, respectively.
By inspecting Table \ref{tab.FC_Bayar} we observe that, when the attack works well against the FC network with $K = N$ (last line of the tables), that is, when the attack targeted to the original BayarNet2016 model can be successfully transferred to the full feature FC detector, the proposed randomization strategy helps and  a significant gain in the Accuracy can be achieved, at the expense of a minor Accuracy reduction in the absence of attacks.
Specifically, the Accuracy gain is about 20-30\% for $K = 30$ and 30-50\% for $K = 10$, while the accuracy reduction in the absence of attacks  is only 2-4\%,
the exact value depending on the task (with the exception of the resizing detection task with BarniNet2018, where a more significant loss of performance is experienced without attacks, see Table \ref{tab.FC_ICIP}).
In some cases, however, it happens that the Accuracy is already large also for $K = N$, i.e., the attack fails against the full feature FC detector, meaning that the attack targeted to the original  CNN BayarNet2016 cannot be transferred to the full feature FC detector. Stated in another way, just re-training the FC network on a different set (a subset of the original training in our case) decreases by itself the attack success rate.
This behavior confirms the findings in \cite{barni2019transferability}, showing that, at least for image forensic applications, the adversarial examples are generally non-transferable, in contrast to what happens in typical pattern recognition applications \cite{PaperTransf16}.

A similar behavior can be observed in Table \ref{tab.FC_ICIP}, where we see that for $K = N$ the attack is even less effective than before (hence it is less transferable).
Then, again, in these cases, the randomization defence is not necessary (e.g. for the case of PGD, the Accuracy with $K=N$ is 89.7\% for resizing and 85.3\% for median filtering). In the other cases, i.e., when the Accuracy with $K=N$ is low, the randomization approach  increases the Accuracy by 20-30\%.
%
%
\begin{table}[htbp]
\scriptsize
\centering
\vspace{-0.2cm}
\caption{Accuracy (\%) of the RDFS detector based on FC network, for the case of BayarNet2016 \cite{bayar2016deep}.}
\label{tab:Rand_FC}
\renewcommand\arraystretch{1.1}
\vspace{0.2cm}
\setlength{\tabcolsep}{3pt}
\begin{tabular}
{a|M{0.4cm}|M{0.4cm}|M{0.45cm}|M{0.45cm}|M{0.4cm}|M{0.4cm}|M{0.45cm}|M{0.45cm}|M{0.4cm}|M{0.4cm}|M{0.45cm}|M{0.45cm}|}
\cline{2-13}
\rowcolor{grannysmithapple}
  \cellcolor{white}
		&  \multicolumn{4}{c|}{Resize} & \multicolumn{4}{c|}{Median Filtering} & \multicolumn{4}{c|}{CL-AHE} \\
\hline
\rowcolor{gray!15}
 \multicolumn{1}{|c|}{\cellcolor{gray!45}{K}}
&  {\tiny  No Attk} & {\tiny \bf PGD}  & {\tiny \bf FGSM} & {\tiny \bf BFGS} &  {\tiny  No Attk} & {\tiny \bf PGD}  & {\tiny \bf FGSM} &   {\tiny \bf BFGS}  &  {\tiny  No Attk} & {\tiny \bf PGD}  & {\tiny \bf FGSM} &   {\tiny \bf BFGS}   \\ \hline
\multicolumn{1}{|c|}{\cellcolor{gray!45}{5}}	&  91.0    &   69.9  &  61.6   &   65.6  &   88.7  &  79.8   &  51.0   &  73.0   &   73.0  &  87.4   &  89.2  &  88.0   \\ \hline
\rowcolor{red!35}
\multicolumn{1}{|c|}{10}	&  95.0   &   68.0  &  55.7  &   62.0  &   93.2  &  80.6   &  44.5   &  67.1   &   78.0  &  88.0   &   89.1  &   78.6   \\ \hline
\rowcolor{red!25}
\multicolumn{1}{|c|}{30}	& 97.0    &  58.5   &  43.4  &   48.8  &   96.8  & 79.7    &   30.8  &  56.1  &   80.1  &   89.5  &  90.7   &    64.7 \\ \hline
\multicolumn{1}{|c|}{\cellcolor{gray!45}{50}}& 97.4    &   52.0  &  35.9  &   40.1  &   97.7  &  80.0   &  24.6   &   53.5  &  80.7   &  90.2   &  91.3   &   56.3  \\ \hline
\multicolumn{1}{|c|}{\cellcolor{gray!45}{200}}	& 97.8   &  31.0   &   13.7  &   17.4  &   98.7  &   77.6  &   10.8  &   44.8  &  81.5   &  91.6   &   94.0  &   42.8  \\ \hline
\multicolumn{1}{|c|}{\cellcolor{gray!45}{400}}& 97.7   &   20.7  &   7.2  &   9.1    &   98.8  &  76.6   &  7.5   &   42.6  &  81.3   &  91.8   &   94.5  &   41.0  \\ \hline
\multicolumn{1}{|c|}{\cellcolor{gray!45}{600}} &  97.9  &   16.4  &  5.4   &   7.1    &   98.9  &  80.5   &  6.0   &   43.0  &  80.5  &  91.5   &  93.8   &    36.6 \\ \hline
\rowcolor{blue!15}
\multicolumn{1}{|c|}{$N$}&  98.0  &  31.5   &  0.6   &  20.3 &   99.0  &   81.9  &   4.3  &    39.7  &  80.5   &   91.8  &   93.9  &   35.1  \\ \hline
	\end{tabular}
\vspace{-0.5cm}
\label{tab.FC_Bayar}
\end{table}

\begin{table}[htbp]
\scriptsize
\centering
\vspace{-0.4cm}
\caption{Accuracy (\%) of the RDFS detector based on FC network, for the case of BarniNet2018  \cite{BCNT18}.}
\vspace{0.2cm}
	\renewcommand\arraystretch{1.1}
\setlength{\tabcolsep}{3pt}
\begin{tabular}
{a|M{0.4cm}|M{0.4cm}|M{0.45cm}|M{0.45cm}|M{0.4cm}|M{0.4cm}|M{0.45cm}|M{0.45cm}|M{0.4cm}|M{0.4cm}|M{0.45cm}|M{0.45cm}|}
\cline{2-13}
\rowcolor{grannysmithapple}
  \cellcolor{white}
		&  \multicolumn{4}{c|}{Resize} & \multicolumn{4}{c|}{Median Filtering} & \multicolumn{4}{c|}{CL-AHE} \\
\hline
\rowcolor{gray!15}
 \multicolumn{1}{|c|}{\cellcolor{gray!45}{K}}
&  {\tiny  No Attk} & {\tiny \bf PGD}  & {\tiny \bf FGSM} & {\tiny \bf BFGS} &  {\tiny  No Attk} & {\tiny \bf PGD}  & {\tiny \bf FGSM} &   {\tiny \bf BFGS}  &  {\tiny  No Attk} & {\tiny \bf PGD}  & {\tiny \bf FGSM} &   {\tiny \bf BFGS}   \\ \hline
\multicolumn{1}{|c|}{\cellcolor{gray!45}{5}}	& 74.4    &  67.7   &  58.7  &   60.7    & 97.2    &   83.3  &  48.3   &  77.1   & 87.4 &   47.2  &   63.7  &  47.0   \\ \hline
\rowcolor{red!35}
\multicolumn{1}{|c|}{10}	& 78.6   &   71.9  &  59.9    &   63.0   &  98.8  &   86.1  &  44.3   &  79.2   &  91.1  &  55.6   &  68.8   &  48.3  \\ \hline
\rowcolor{red!25}
\multicolumn{1}{|c|}{30}	& 92.7  &  81.8   &  65.5   &    70.7  &   99.4  &  88.5   &  30.0   &  79.6   &  94.3  &  56.7   &  76.3   &  39.8  \\ \hline
\multicolumn{1}{|c|}{\cellcolor{gray!45}{50}} & 96.8  &  85.2   &  66.8   &   73.0   &   99.6  &  87.4   &  21.9   &   76.6  &  95.1  &   50.6  &  80.0   &   35.3  \\ \hline
\multicolumn{1}{|c|}{\cellcolor{gray!45}{200}}	& 99.7   &  88.0   &  69.6   &    77.9   &  99.6   &  88.6  &  17.0   &  76.2   &  96.9   &  48.5   &  83.0   &   26.0  \\ \hline
\multicolumn{1}{|c|}{\cellcolor{gray!45}{400}} & 99.8  &   89.3  &   71.8   &   80.0   &   99.6  &  88.1   &  15.6   &  75.6   &  97.1   &  30.1   &  83.6   &   21.0  \\ \hline
\rowcolor{blue!15}
\multicolumn{1}{|c|}{$N$} &  100   &   89.8  &  75.2  &    81.2   &  99.7   &  85.2   &  13.7   &  71.3   &  98.2   &  33.5   &   34.0  &   26.2  \\ \hline
	\end{tabular}
\vspace{-0.3cm}
\label{tab.FC_ICIP}
\end{table}

\subsection{Results of SVM-based classification}

The results we have got for this case are reported in Table \ref{tab.SVM_Bayar} and \ref{tab.SVM_ICIP} for the case of BayarNet2019  and BarniNet2018, respectively.

In this case, expectedly, the mismatch in the architecture decreases further the success rate of the attack against the full feature SVM detector (case with $K =N$), i.e. it increases the Accuracy, without even resorting to randomization. However, when this is not the case, the randomization helps:
for instance, for BayarNet2016 under the BFGS attack,   the Accuracy passes from 39.4 (for the resizing task), 5.0 (for the median filtering task) and 38.6 (for the CL-AHE task), to 69.6\%, 50.7\%, and 70.5 \% respectively, with a performance loss without attacks  of 2.2\% in the Accuracy.
Using less features, e.g $K = 10$, the Accuracy against the attacks can be improved further in general, though at the expense of a higher loss of performance without attacks.
\begin{table}[htbp]
\scriptsize
\centering
\vspace{-0.3cm}
\caption{Accuracy (\%) of the RDFS detector based on SVM, for the case of  BayarNet2016 \cite{bayar2016deep}.}
\vspace{0.2cm}
\renewcommand\arraystretch{1.1}
\setlength{\tabcolsep}{3pt}
\begin{tabular}
{a|M{0.4cm}|M{0.4cm}|M{0.45cm}|M{0.45cm}|M{0.4cm}|M{0.4cm}|M{0.45cm}|M{0.45cm}|M{0.4cm}|M{0.4cm}|M{0.45cm}|M{0.45cm}|}
\cline{2-13}
\rowcolor{grannysmithapple}
  \cellcolor{white}
		&  \multicolumn{4}{c|}{Resize} & \multicolumn{4}{c|}{Median Filtering} & \multicolumn{4}{c|}{CL-AHE} \\
\hline
\rowcolor{gray!15}
 \multicolumn{1}{|c|}{\cellcolor{gray!45}{K}}
&  {\tiny  No Attk} & {\tiny \bf PGD}  & {\tiny \bf FGSM} & {\tiny \bf BFGS} &  {\tiny  No Attk} & {\tiny \bf PGD}  & {\tiny \bf FGSM} &   {\tiny \bf BFGS}  &  {\tiny  No Attk} & {\tiny \bf PGD}  & {\tiny \bf FGSM} &   {\tiny \bf BFGS}   \\ \hline
\multicolumn{1}{|c|}{\cellcolor{gray!45}{5}}	 & 79.6    &  59.0   &  58.0   &   58.7   &  80.3  &   69.8  &  47.5  &  66.1    &  74.4   &  90.7   &  89.2  &  87.3   \\ \hline
\rowcolor{red!35}
\multicolumn{1}{|c|}{10}	&  87.0   &  60.5   &  58.9   &   59.9   &  87.6  &  70.8   &  33.8  &  63.2   &  80.4   &  90.7   &  90.4  &  81.5   \\ \hline
\rowcolor{red!25}
\multicolumn{1}{|c|}{30}	&  92.8   &  70.9   &  70.1   &  69.6   &   94.5  &  63.3   &  19.1   &  50.7   &  80.5  &   89.6  &  90.9   &   70.5  \\ \hline
\multicolumn{1}{|c|}{\cellcolor{gray!45}{50}} &   94.3  &  75.5   &  75.6   &  75.0   &   96.2  &  66.8   &  13.1   &  42.0   &  80.7   &  89.8   & 91.0   &  62.3   \\ \hline
\multicolumn{1}{|c|}{\cellcolor{gray!45}{200}}	&   95.5  &  65.0  &  63.9  &  64.2  &   97.7  &   57.2  &   3.8  &   22.1  &  80.0   &   91.0  &  93.4   &  43.7   \\ \hline
\multicolumn{1}{|c|}{\cellcolor{gray!45}{400}}&  94.8   &  43.4   &  66.4  &  28.1   &   98.0  &  50.3   &  1.9   &  14.9   &  79.7  &   91.2  &  93.8   &   39.7  \\ \hline
\multicolumn{1}{|c|}{\cellcolor{gray!45}{600}} &   95.4  &  47.9   &  25.2   &  32.3   &  98.1   & 45.0   &  1.3   &  11.0   &  79.4   &  91.3   &  94.2  &  40.1  \\ \hline
\rowcolor{blue!15}
\multicolumn{1}{|c|}{$N$}& 95.1    &  58.4   & 31.0    &  39.4  &   98.0  &   29.6  &   0.6  &   5.0  &   79.5  &  91.8   &   95.0  &   38.6   \\ \hline
	\end{tabular}
\vspace{-0.5cm}
\label{tab.SVM_Bayar}
\end{table}
\begin{table}[htbp]
\scriptsize
\centering
\vspace{-0.3cm}
\caption{Accuracy (\%) of the RDFS detector based on SVM, for the case of BarniNet2018 \cite{BCNT18}.}
\vspace{0.2cm}
\renewcommand\arraystretch{1.1}
\setlength{\tabcolsep}{3pt}
\begin{tabular}
{a|M{0.4cm}|M{0.4cm}|M{0.45cm}|M{0.45cm}|M{0.4cm}|M{0.4cm}|M{0.45cm}|M{0.45cm}|M{0.4cm}|M{0.4cm}|M{0.45cm}|M{0.45cm}|}
\cline{2-13}
\rowcolor{grannysmithapple}
  \cellcolor{white}
		&  \multicolumn{4}{c|}{Resize} & \multicolumn{4}{c|}{Median Filtering} & \multicolumn{4}{c|}{CL-AHE} \\
\hline
\rowcolor{gray!15}
 \multicolumn{1}{|c|}{\cellcolor{gray!45}{K}}
&  {\tiny  No Attk} & {\tiny \bf PGD}  & {\tiny \bf FGSM} & {\tiny \bf BFGS} &  {\tiny  No Attk} & {\tiny \bf PGD}  & {\tiny \bf FGSM} &   {\tiny \bf BFGS}  &  {\tiny  No Attk} & {\tiny \bf PGD}  & {\tiny \bf FGSM} &   {\tiny \bf BFGS}   \\ \hline
\multicolumn{1}{|c|}{\cellcolor{gray!45}{5}}	&  74.8     &  73.6   &  65.4   &   66.4  &  97.2   &   83.1  &  46.3   &  78.1   &  88.3   &   59.6  & 64.8  &  50.7   \\ \hline
\rowcolor{red!35}
\multicolumn{1}{|c|}{10}	&  82.7    &  78.1   &   68.0  &   69.2  &   98.3  &   85.7  &  42.0  &  80.3   &  91.2   &  67.0   &   74.5  &  58.0   \\ \hline
\rowcolor{red!25}
\multicolumn{1}{|c|}{30}	&  95.1    &  86.1   &  72.8   &   76.0  &   99.3  &  86.1   &   25.8  &   77.2  &  93.4   &  63.4   &  86.5   &  54.8   \\ \hline
\multicolumn{1}{|c|}{\cellcolor{gray!45}{50}}&  97.5    &  88.2  &  73.2   &   77.2  &  99.3   &   84.1  &   18.7  &   74.0  &  94.5   &   56.7  &   90.3  &  46.7   \\ \hline
\multicolumn{1}{|c|}{\cellcolor{gray!45}{200}}& 99.6    &   88.5  & 68.1    &   75.7  &  99.6   &   88.2  &  15.9   &   75.2  &  96.4   &  36.4   & 94.0    &  24.7   \\ \hline
\multicolumn{1}{|c|}{\cellcolor{gray!45}{400}}&  99.7   &  90.0   &  67.6   &   77.7   &  99.6   &  86.6   &  13.5   &  70.1   &  97.0   &  26.1   &   94.1  &   16.1  \\ \hline
\rowcolor{blue!15}
\multicolumn{1}{|c|}{$N$}&  99.8   & 90.6    & 66.2   &   83.8   &  99.7   &  86.4   &  12.0   &  69.8   & 97.3   &  22.3   &  94.6   &  11.0   \\ \hline
	\end{tabular}
\label{tab.SVM_ICIP}
\end{table}


\enlargethispage{\baselineskip}

\vspace{-0.3cm}

\section{CONCLUDING REMARKS}
\label{sec.CONC}
\vspace{-0.1cm}

Following \cite{chen2019secure}, we have evaluated the feasibility of using deep feature randomization to improve the robustness of CNN detectors against adversarial examples by hindering the transferability of the attacks. Our experiments carried out in a wide variety of scenarios reveal that feature randomization somewhat helps in decreasing the transferability of the attacks, hence improving the security of the detection, even if, in some cases, the mismatch in the architecture between the original CNN and the new detector is enough to prevent the transferability.
%
Future work will focus on increasing the strength of the attacks to improve the transferability of the adversarial examples.
Given the complexity of the decision boundary learnt by the CNNs, this is not an easy task.
The amount of distortion introduced in the image by the attack, or the value assumed by the decision function, in fact, represent only an inaccurate proxy for the attack strength, since controlling the amount of distortion (e.g., by letting the gradient descent continue until a limit PSNR is reached for the attack),
or setting a safe margin on the decision function for the attack, do not necessarily result in a stronger attack.
%
Another direction for future research is to investigate a scenario more favourable to the attacker, where the attacker is aware of the randomization-based defence.
In particular, we can assume that the attacker is aware of the
feature selection mechanism and the architecture of the detector (only the secret key is unknown), and then can run a more powerful attack, for instance by targeting an
expected version of the classifier (in a way that resembles the Expectation over Transformation (EOT) attack  \cite{athalye2018obfuscated}).
From the defender's side, one could try to improve the effectiveness of the RDFS scheme by performing feature regularization during the training of the original CNN, in such a way to reduce the gap with the theoretical analysis carried out in \cite{chen2019secure} about the effectiveness of the feature selection strategy.

%


\bibliographystyle{IEEEbib}
\bibliography{ICASSP20}

\end{document}